\documentclass[runningheads]{llncs}

\usepackage{eccv}

\usepackage{eccvabbrv}

\usepackage{graphicx}
\usepackage{booktabs}

\usepackage[accsupp]{axessibility}  % Improves PDF readability for those with disabilities.

\usepackage{hyperref}

\usepackage{orcidlink}

\begin{document}

\title{Patterns of Creativity: How User Input Shapes AI-Generated Visual Diversity}

\titlerunning{ }

\author{Maria-Teresa De Rosa Palmini\orcidlink{0009-0005-7700-0097} \and
Eva Cetinic\orcidlink{0000-0002-5330-1259}}

\authorrunning{De Rosa Palmini and Cetinic}

\institute{University of Zurich, Zurich, Switzerland\\
\email{maria-teresa.derosa-palmini@uzh.ch}\\
\email{eva.cetinic@uzh.ch}
}

\maketitle

\begin{abstract}
Recent critiques of Artificial-intelligence (AI)-generated visual content highlight concerns about the erosion of artistic originality, as these systems often replicate patterns from their training datasets, leading to significant uniformity and reduced diversity.  Our research adopts a novel approach by focusing on user behavior during interactions with Text-to-Image models. Instead of solely analyzing training data patterns, we examine how users’ tendencies to create original prompts or rely on common templates influence content homogenization. We developed three originality metrics—lexical, thematic, and word-sequence originality—and applied them to user-generated prompts from two datasets, DiffusionDB and Civiverse. Additionally, we explored how characteristics such as topic choice, language originality, and the presence of NSFW content affect image popularity, using a linear regression model to predict user engagement. Our research enhances the discourse on AI’s impact on creativity by emphasizing the critical role of user behavior in shaping the diversity of AI-generated visual content. 

\end{abstract}

\section{Introduction}

The latest advancements in Computer Vision and Natural Language Processing have paved the way for numerous applications in generative models, particularly in text-guided image generation. Innovations like DALL-E 2 \cite{ramesh2022hierarchical} and Stable Diffusion \cite{rombach2022high} have built upon advanced techniques in combined image and text embedding learning, such as Contrastive Language and Image Pre-training (CLIP) \cite{radford2021learning}, enabling the generation of photorealistic and aesthetically appealing images from textual prompts. The applications of generative Artificial-intelligence (AI) Text-to-Image (TTI) models are currently utilized in diverse sectors, such as news media graphics \cite{liu2022opal} and product design \cite{ko2023large}. However, despite these advancements, concerns are growing about their heavy reliance on training data, which often leads to the perpetuation of Western-centric biases related to culture, race, and gender \cite{bianchi2023easily}, \cite{cho2023dall}, as well as a lack of authenticity in the generated content, resulting in the emergence of a visually homogenous culture.

While much of the existing research has focused on algorithmic biases and dataset limitations in TTI models \cite{chauhan2024identifying}, \cite{luccioni2024stable}, \cite{shankman2024staring} less attention has been given to the role of human agency, particularly through the user-provided text descriptions, also known as "prompts", in shaping the AI-generated visual culture. Research in prompt engineering has proposed various guidelines for structuring prompts, recommending specific language and techniques to optimize image quality by offering effective strategies for prompt formulation \cite{liu2022design}, \cite{oppenlaender2023taxonomy}, \cite{wang2023reprompt}. Online communities, such as the subreddit \texttt{r/StableDiffusion}\footnote{\url{https://www.reddit.com/r/StableDiffusion/}}, and resources like \textit{A Beginner’s Guide to Prompt Design for Text-to-Image Generative Models}\footnote{\url{https://towardsdatascience.com/a-beginners-guide-to-prompt-design-for-text-to-image-generative-models-8242e1361580}} and \textit{The Art and Science of Prompt Engineering}\footnote{\url{https://medium.com/@alimelki/the-art-and-science-of-prompt-engineering-mastering-ai-communication-cc55261bfe24}}  provide users with prompt examples and strategies, helping them achieve impressive results with minimal trial and error. However, this standardized prompting process, while effective for creating visually appealing images, may also lead to the repetitive use of similar language structures, reinforcing a culture of visual uniformity. Additionally, the persistent use of a coding-syntax-like language in prompts may have established a fixed linguistic template, constraining users to narrow parameters and limiting their ability to explore and experiment with a broader and more diverse range of linguistic expressions and thematic subjects.

This study aims to bridge the gap in understanding how user-generated prompts influence the originality of AI-generated visual content. It suggests that uniformity in prompts, both in terms of topics and linguistic choices, may be one of the factors contributing to visual homogenization, potentially limiting stylistic diversity in AI outputs while also perpetuating certain cultural biases. To do so, we focus not only on the topics users prompt about, but also on the originality of the language choices used in these prompts. To achieve this, we developed three originality metrics: \textit{lexical originality} (the uniqueness of words), \textit{thematic originality} (the diversity of topics), and \textit{word-sequence originality} (the novelty of word combinations). These metrics are based on the concept of "novelty" defined by Shah et al. \cite{shah2003metrics}, which considers an idea novel if it significantly deviates from existing ones, as determined by the rarity of its attributes compared to a predefined set or across a collection of ideas. In our study, we apply this concept to analyze the frequency of lexical choices, thematic elements, and word sequences within two extensive prompt datasets, \textbf{DiffusionDB} \cite{wang2022diffusiondb} and \textbf{Civiverse} \cite{palmini2024civiverse}, identifying as original those that deviate significantly from the dataset norm.

In addition to assessing prompt originality, we aimed to identify which other prompt features influence the popularity of AI-generated images on open-source platforms like CivitAI \cite{civitai}, as the type of content that resonates with users often reflects broader aesthetic trends in emerging AI-generated visual culture. To achieve this, we developed a Linear Regression model to predict user engagement based on various prompt characteristics, including originality metrics, specific topics identified through topic modeling, and the presence of NSFW content. By analyzing the impact of these features on image popularity, we sought to determine which elements are most likely to shape aesthetic preferences in AI-generated content. This analysis provides insight into whether popular content in open-source platforms contributes to visual homogenization and bias reinforcement, or promotes greater diversity in AI outputs.

By examining the interplay between user behavior and AI outputs, this research addresses a critical gap in understanding how human agency influences the cultural products of AI systems. It contributes to the field of human-computer interaction (HCI) by highlighting the social and creative dynamics that may drive visual uniformity in AI-generated content. Ultimately, this study offers new insights into how human-AI collaboration can either foster diversity in visual outputs or reinforce existing biases and patterns, both artistic and cultural. These findings underscore the need for platform developers and online communities to revisit current guidelines, fostering a creative environment that encourages exploration, innovation, and the representation of a wider range of perspectives in AI-generated visual culture.

\section{Theoretical Background and Related Work}

The emergence of generative AI technologies, particularly TTI models, has transformed creative processes, enabling users to generate visuals based on textual descriptions. While these tools open new avenues for artistic expression, they also raise important questions about how technology and human agency interact in shaping the creative outcomes. In this section, we explore key debates on technological determinism, the critical role of human agency, and how prompt engineering practices might impact user creativity. By examining these issues, we aim to contextualize the relationship between user-generated prompts and originality of AI-generated outputs.

\section {The Notion of Technological Determinism}
A central debate surrounding technology’s role in society is framed by the theory of technological determinism, which asserts that technology is the primary force driving social, cultural, and artistic change \cite{drew2016technological}. However, this view has been critiqued for oversimplifying the intricate relationship between humans and technology, ignoring the broader context in which technology operates \cite{de2016organizing}. Critics argue that technology is not an autonomous force but is shaped by an array of social, cultural, and political factors that influence its development and use. As a result, its effects are not solely determined by technological capabilities but are also the product of human decisions, values, and the specific contexts in which it is applied \cite{tessema2021technological}.

\subsection{The Role of Human Agency}

Given the critique of technological determinism, the role of human agency becomes crucial in understanding how AI technologies influence creative outputs. While generative AI has been praised for its collaborative potential, particularly in ideation and art-making, recent studies have shown that human-AI interaction remains central to producing meaningful creative outcomes \cite{cetinic2022understanding}, \cite{chiou2023designing}, \cite{grba2024art}, \cite{paananen2023using}. Models of human-AI collaboration range from systems with elevated computational awareness \cite{davis2015enactive} to frameworks that emphasize real-time, bidirectional communication \cite{mccormack2019silent}, \cite{thelle2021spire}. Additionally, mixed-initiative approaches, where both human and AI agents contribute to the creative outcome, have been shown to effectively support creative processes \cite{lin2023beyond}, \cite{zhu2018explainable}.

Despite these advancements, there is an increasing need to reestablish human agency as central to the creative process, particularly as the reliance on Large Language Models (LLMs) like ChatGPT \cite{OpenAI2023} and other generative technologies grows significantly. While these tools offer new possibilities for artistic expression, they also introduce significant risks, most notably the potential erosion of creativity through the repetitive generation of formulaic, uninspired outputs. Critical AI studies have emphasized that TTI generated images often lack a clear authorial figure, raising complex questions about the notion of authorship in the context of AI-generated art. Specifically, some scholars suggest that the concept of authorship is being displaced, with the role of the prompt-maker being minimized or overlooked entirely in favor of assigning creative identity to the scientists and programmers behind the models themselves \cite{wilde2023generative}. This debate revisits older disputes from the field of computational art, where advanced technologies often led to a blurring of boundaries between human creative decision-making and the highly formalized processes that seek to emulate it \cite{grba2023renegade}. Additioanly, the depiction of AI systems as "creative collaborators" has amplified these uncertainties, reinforcing misconceptions about the concept of machine agency and concealing the genuine relationship between human creators and AI-generated processes \cite[pp.27–28, 241–43]{audry2021art}, \cite[pp.3–5]{grba2022deep}.

In this context, the risk of visual homogenization is particularly evident in TTI models, where users often rely on standardized prompts and familiar visual styles, resulting in repetitive and uninspired imagery. This issue reflects broader challenges in the design of Creativity Support Tools (CSTs), which, while one the hand intend to foster and enhance creativity, can impose rigid structures that limit creative exploration and diversity. To mitigate this, maintaining active human engagement in AI-assisted creative processes is crucial, as it ensures the creation of original and meaningful content while also enhancing the effectiveness of CSTs \cite{peschl2024human}. Furthermore, research in both Creativity Research (CR) and Human-Computer Interaction (HCI) emphasizes the importance of promoting divergent thinking and encouraging exploratory practices, which are often overshadowed with the role of creativity being constrained by restrictive and technology-guided frameworks \cite{frich2018hci}.

\subsection{Prompt Engineering and Its Impact on Originality}

The rise of generative models has highlighted the importance of prompt design, resulting in numerous techniques being developed for prompt engineering. The concept gained prominence from a well-known post by Gwern Branwen \cite{branwen2020gpt3} about GPT-3’s ability to generate creative fiction, where it was suggested that mastering prompt design could become a new mode of interacting with models. Users would simply need to craft prompts in a way that draws out the required information and abstractions for completing specific tasks. As the field has evolved, prompt engineering has greatly improved the capabilities of generative models, particularly in TTI systems.

Many TTI platforms now offer guidelines to help users create prompts that enhance image quality. For instance, in earlier iterations of Stable Diffusion, phrases like \textit{cinematic}, \textit{highly detailed}, and \textit{8k} were essential for generating high-quality images, though recent advancements in the model have made these terms less critical. Similarly, OpenAI's DALL-E offers guidance for users to improve image generation\footnote{\url{https://platform.openai.com/docs/guides/prompt-engineering/}}, recommending the inclusion of specific details like elements, settings, artistic styles, or emotions to produce more accurate and refined outputs.

In addition to prompting guidelines, several techniques have been developed to optimize prompt creation and improve output quality. Methods like few-shot prompting \cite{brown2020language}, \cite{zhao2021calibrate} and chain-of-thought prompting \cite{wei2022chain}, \cite{zhao2021calibrate} help guide models by providing limited examples or by breaking down tasks into logical steps. Research has also emphasized the importance of style keywords and modifiers for generating high-quality images \cite{liu2022design}, \cite{oppenlaender2023taxonomy}. In more, recent advancements in the field include automated prompt generation techniques with reinforcement learning-based tools like \textit{Promptist} \cite{hao2024optimizing}, which enhance prompt effectiveness across different models. These automated methods are further supported by interactive tools like \textit{Promptify} \cite{brade2023promptify} or \textit{PromptCharm} \cite{wang2024promptcharm}, which refine prompts iteratively based on user feedback.

Although these aforementioned developments have increased the sophistication of prompt engineering, this increasing emphasis on prompt optimization has inadvertently led to the development of standardized "prompt templates," where users tend to rely on familiar structures and themes, contributing to a growing visual uniformity in AI-generated content. Research into the thematic trends of user-generated prompts on platforms like \textit{Midjourney} has shown that popular styles and surface-level aesthetics—such as fantasy, game art, and anime—are often prioritized over deeper artistic elements like narrative and authenticity \cite{mccormack2024no}, \cite{sanchez2023examining}. This tendency has resulted in a cultural uniformity in the topics generated, reinforcing stylistic norms and leading to the emergence of predominantly homogenized outputs. However, while much of the existing research has focused on what users are prompting about, there has been limited attention to the linguistic diversity in how these prompts are crafted. Exploring therefore language originality—through patterns in lexical choices, thematic elements, and word combinations—is crucial in understanding how prompt engineering might unintentionally limit visual diversity. Addressing this gap can reveal whether current prompt practices and guidelines are contributing to or counteracting the trend toward homogenization in the AI-generated visual culture.

\section{Methodology}
In this section, the approach to evaluating the impact of prompt originality on the diversity of AI-generated content is outlined. \textit{Lexical}, \textit{word-sequence}, and \textit{thematic originality} are analyzed across two datasets to assess the diversity of user-generated prompts between them. Furthermore, the correlations between prompt characteristics and user engagement are also explored, providing insights into how human input is shaping creative outcomes and popular trends in AI-generated visuals.

\subsection{Datasets Overview}
In this study, prompt originality is examined through the use of two large-scale datasets: DiffusionDB and Civiverse. These datasets capture user interactions with TTI models but differ significantly in their origins, purposes, and content regulations. DiffusionDB comes from a strictly moderated environment, while Civiverse represents a more open platform with fewer restrictions. This variation is essential for understanding whether different content guidelines might influence not only the linguistic originality in user-generated prompts but also the diversity of the resulting AI-generated visuals.

\paragraph{DiffusionDB}
The DiffusionDB dataset \cite{wang2022diffusiondb} was collected over a two-week period, from October 6 to October 20, 2022, by gathering user-generated images and their associated metadata, such as seed values, step counts, CFG scale, and image size, from the official Stable Diffusion Discord server. With approximately 1.8 million unique text prompts, primarily contributed by experienced users and early adopters of Stable Diffusion, the dataset may exhibit a bias toward more advanced prompting techniques, as noted by the dataset creators. As Stable Diffusion is a widely-used open-source, large-scale TTI model, DiffusionDB is released under a CC0 1.0 license, allowing unrestricted public use of the content \cite{stabilityai2022b}. The Discord server enforces strict moderation policies, prohibiting illegal, hateful, NSFW, and personally identifiable content \cite{stabilityai2022a}, ensuring that the dataset remains focused on safe-for-work, creative outputs.

\paragraph{Civiverse}

The Civiverse dataset \cite{palmini2024civiverse} is sourced from the CivitAI platform \cite{civitai}, which facilitates image generation and the exchange of Stable Diffusion model derivatives. The platform has experienced rapid growth, with a substantial volume of user-generated content uploaded daily. The dataset, assembled from this platform, contains metadata for over 6.5 million images posted between October 2023 and April 2024, such as the web URL, image hash, post ID, timestamp, anonymized user data, model information, and content rating. It is worth noting that CivitAI imposes fewer content restrictions, which is reflected in the significant proportion of NSFW content on the platform. To be more specific, the percentage of content rated above PG-13 increased from 55.19\% in October 2023 to 72.94\% in April 2024, highlighting the platform's more permissive content policies.

\subsection{Topic Modeling}

Inspired by previous work in prompt analysis \cite{mccormack2024no}, \cite{sanchez2023examining}, the thematic coverage of both the DiffusionDB and Civiverse datasets was first explored. This was considered essential not only for comparison but also as a critical step in developing the \textit{thematic originality} metric (\ref{sec:thematic_originality}), as the thematic clusters and their associated keywords from both datasets would be used as labels for the corresponding prompts. Given the disparity in dataset sizes—1.8 million prompts from DiffusionDB and over 6 million from Civiverse—a direct comparison was deemed non-representative. To address this, a random sample of 1.8 million prompts from Civiverse was taken to ensure a balanced analysis.

Due to the variability in raw prompt formats, individual prompt specifiers, text fragments specifying desired image characteristics (e.g., "highly detailed," "cinematic"), were analyzed. Following a method similar to Sanchez \cite{sanchez2023examining}, prompts were segmented by commas, and specifiers that appeared at least 300 times and were under 35 characters in length were selected. This process resulted in approximately 11,000 specifiers for DiffusionDB and 19,000 for Civiverse prompts.

Additionally, the MiniLMv2 model \cite{wang2020minilmv2} was used to transform the specifiers into 384-dimensional embeddings. For topic modeling and clustering, BERTopic was employed, utilizing UMAP \cite{mcinnes2018umap} for dimensionality reduction and HDBSCAN \cite{campello2013density} for clustering. Additionally, class-specific term frequency-inverse document frequency (c-TF-IDF) was further applied to extract key terms within clusters, facilitating the manual labeling of thematic groups.

\subsection{Prompt Originality Metrics}

To comprehensively assess the diversity of user-generated prompts in AI-generated visual content, three originality metrics, \textit{lexical originality}, \textit{word-sequence originality}, and \textit{thematic originality}, were calculated. These metrics draw on the concept of "novelty" as outlined by Shah et al. \cite{shah2003metrics}, which defines an idea as novel if it deviates considerably from existing ones, measured by the rarity of its characteristics in relation to a predefined set or a broader collection of ideas. Each metric captures a different dimension of prompt originality by evaluating the uniqueness of individual words, the contextual relationships between words, and the novelty of thematic combinations, respectively. By applying these metrics, the goal was to quantify the level of novelty introduced by the prompts in the two datasets, providing valuable insights into the originality of user interactions with TTI models.

Before calculating the originality scores, all text prompts underwent preprocessing steps, including stopword and special character removal, as well conversion to lowercase to ensure consistency in analysis.

\subsubsection{Lexical Originality}

To assess the \textit{lexical originality} of user-generated prompts, a metric that evaluates the rarity of individual words within each prompt relative to the entire dataset, is introduced. This measure plays a crucial role in distinguishing prompts that rely on common words from those that use more distinctive vocabulary, offering insights into how creatively users are interacting with TTI models.

The process begins by calculating the frequency of each word across the dataset to determine how common or rare it is. The rarity score for each word is then computed as the inverse of its frequency, meaning that frequently appearing words receive a low rarity score, while rarely used words are assigned a higher score. To ensure stability in the calculation and prevent division by zero, a small constant \( \epsilon \) is added to the frequency, allowing the formula to function smoothly even for extremely common words.

To discourage excessive repetition within a prompt, a penalty is applied based on how often a word is repeated. The more frequently a word appears in the same prompt, the higher the penalty:

\[
\text{Penalty}_{\text{rep}} = \frac{\sum_{i=1}^{n} (\text{count}(w_i) - 1)}{n}
\]

where \( n \) is the total number of words in the prompt, and \( \text{count}(w_i) \) represents how many times word \( w_i \) appears. In addition to the repetition penalty, another penalty is imposed for prompts that rely heavily on the most frequently used words across the entire dataset. These commonly used words further reduce the originality score, as they indicate less creative variation.

To ensure that longer prompts do not automatically receive higher scores, the \textit{lexical originality} score is adjusted based on the length of the prompt. This adjustment is proportional to the ratio of the number of words in the prompt to the maximum number of words in any prompt within the dataset:

\[
S_{\text{adjusted}} = S_{\text{original}} \times \frac{\text{num\_words}}{\text{max\_words}}
\]

The final \textit{lexical originality} score is then determined by summing the rarity scores of all the words, applying both the repetition and common word penalties, and adjusting for the prompt’s length:

\[
S_{\text{final}} = \max\left(S_{\text{adjusted}} - \text{Penalty}_{\text{rep}} - \text{Penalty}_{\text{common}}, 0\right)
\]

\subsubsection{Word-Sequence Originality}

To address the limitations of \textit{lexical originality}, an additional metric, that is \textit{word-sequence originality}, is introduced. While \textit{lexical originality} measures the rarity of individual words within a dataset, it might still overlook how those words are combined in a broader context. As a result, a prompt might contain rare words but still follow conventional word pairings, something that can be limiting its creative potential. By focusing on the probabilities of transitions between words, \textit{word-sequence originality} captures the contextual relationships, assessing how likely one word is to follow another and providing a deeper layer of originality evaluation in word sequences.

For example, in a TTI prompt like \textit{"beautiful sunset"}, the transition from \textit{"beautiful"} to \textit{"sunset"} is common, contributing less to the sequence originality score. In contrast, a phrase like \textit{"sunset underwater"} may be rare, even if \textit{"sunset"} and \textit{"underwater"} are common individually. Since the transition between these two words is unexpected, it results in a higher sequence originality score.

The sequence originality of each prompt was calculated using a Markov Chain-based model \cite{tolver2016introduction}. First, each prompt was tokenized into words, stopwords are removed, and bigrams (pairs of consecutive words) were generated. The transition probability \( P(w_2 \mid w_1) \), representing the likelihood of \( w_2 \) following \( w_1 \), was calculated as:

\[
P(w_2 \mid w_1) = \frac{\text{count}(w_1, w_2)}{\sum_{w'} \text{count}(w_1, w')}
\]

The sequence originality score \( S \) for each prompt was then calculated by summing the negative logarithms of these transition probabilities:

\[
S_{\text{seq}} = -\sum_{i=1}^{n-1} \log P(w_{i+1} \mid w_i)
\]

Rare transitions (low probability ones) contribute more to the score, while common transitions contribute less. The score is then normalized by dividing the total by the number of bigrams to ensure comparability across prompts of varying lengths. If a bigram is not found in the dataset, a small probability of \( 1 \times 10^{-5} \) is assigned to prevent zero probabilities, ensuring the calculation remains unbiased.

\subsubsection{Thematic Originality}
\label{sec:thematic_originality}

To evaluate the novelty of topics within prompts, a metric called \textit{thematic originality} is developed. This metric measures how unique a set of topics within a prompt is relative to the overall dataset by analyzing both individual topic frequency and the co-occurrence of topics. This is essential for understanding the diversity of user-generated content and determining whether prompts follow repetitive patterns or introduce more original thematic combinations.

The calculation of \textit{thematic originality} begins with the assignment of thematic labels to each prompt based on the results of topic modeling. For instance, a label like \textit{High Resolution} may be associated with keywords such as "8k", "ultra detailed", "4k" and "high definition." These labels are then analyzed for their frequency, both individually and in combination with others. For example, while \textit{Female Subjects} may be a frequent label, its combination with less common ones may still indicate novelty. The frequency of individual labels and their combinations is then calculated, capturing both standalone occurrences and relationships.

The \textit{thematic originality} score for a prompt is calculated by summing the negative logarithms of the probabilities of its topics and their combinations, as follows:

\[
S_{\text{topic}} = -\sum_{i=1}^{m} \log P(t_i) - \sum_{i=1}^{m-1} \sum_{j=i+1}^{m} \log P(t_i, t_j)
\]

Here, \( P(t_i) \) represents the probability of the \(i\)th topic in the dataset, and \( P(t_i, t_j) \) is the probability of the combination of topics \(i\) and \(j\). The negative logarithm is used because it increases the weight of rarer topics and combinations, making them contribute more to the originality score. Conversely, common topics lead to smaller values, lowering the score. For example, if a prompt contains the labels \textit{"Urban Development"} and \textit{"Sustainable Design"}, and these labels, individually and together, appear infrequently in the dataset, the prompt will receive a higher originality score. In contrast, if the topics are common, the score will be lower.

\paragraph{Clustering and Categorization}
To evaluate the originality of prompts across the three metrics introduced, K-Means clustering is used to group the prompts into three categories: Low, Moderate, and High Originality. Instead of applying predefined originality thresholds, which may not generalize well across datasets with different characteristics, K-Means dynamically adjusts to the dataset by clustering based on the centroids of the originality scores. This approach ensures that the method remains flexible and adaptable, regardless of the specific properties of each dataset.

\subsection {Prediction of User Engagement}

In the Civiverse dataset, \textit{LikeCount} serves as a key metadata element, capturing the number of likes each generated image receives on the CivitAI platform. Users typically award \textit{likes} to images that resonate with them on aesthetic, humorous, or intellectual levels, making the number of likes not only a reflection of an image's appeal but also an indicator of how effectively its prompt engages users, thus providing a crucial measure of a prompt’s success within the platform. Consequently, understanding whether specific prompt characteristics can predict \textit{user engagement} becomes essential, as it not only highlights the types of content garnering the most attention but also reveals which images are more frequently displayed and circulated across open-source platforms. As these images gain prominence and visibility, they play a significant role in shaping the overall visual culture of the platform, thereby influencing the dominant aesthetics in AI-generated content and potentially setting trends within the community.

To investigate the relationship between prompt characteristics and user engagement, a subset of approximately 200,000 images and their corresponding metadata from the Civiverse dataset is utilized, given that the DiffusionDB dataset does not include image popularity scores. From this dataset, a balanced sample of images, representing both high and low engagement levels, measured by whether the like count falls above or below the average threshold, is selected, following a similar approach to that of Arazzi et al. \cite{arazzi2023importance}, who predicted \textit{user engagement} in Twitter-related posts. 

In order to determine which prompt characteristics exert the most influence on user engagement, an model based on Linear Regression was developed, as \textit{LikeCount} is a continuous variable \cite[pp. 245–246]{mohri2018foundations}. Each prompt was represented as a vector of features, incorporating various prompt-related characteristics, including the previously calculated \textit{originality metrics} (lexical, word-sequence, and thematic originality), \textit{thematic elements} identified through topic modeling, and the presence of \textit{NSFW content}. After conducting topic modeling on the Civiverse dataset, approximately 70 distinct topics were identified, each associated with specific keywords and manually named based on these keywords. These topics were then utilized as features in the predictive model, with binary indicators assigned for each topic, marked as 1 if any of the topic’s keywords were present in the prompt, and 0 otherwise. Additionally, the presence of NSFW content, already included in the metadata of the Civiverse dataset, was represented as a binary feature. A 5-fold cross-validation procedure was then performed, with 80\% of the data used for training and 20\% for testing, to ensure the robustness of the model. During this process, hyperparameters such as regularization strength and feature selection criteria are fine-tuned to optimize performance while minimizing overfitting. A regularization term of 0.1 is set to control the model’s complexity, ensuring that the coefficients do not become excessively large and that the model captures meaningful patterns rather than fitting to noise. Additionally, a feature selection threshold of 0.05 was applied to focus on statistically significant features, in order for the model to rely on the most relevant predictors.

\section{Results}

\subsection{Civiverse vs DiffusionDB}

\subsubsection{Thematic Similarities and Differences}

The comparative analysis of the Civiverse and DiffusionDB datasets reveals distinct thematic emphases, reflecting the different user bases and content policies of these platforms. Although equal dataset sizes were sampled for this study, topic modeling identified approximately 35 topics for DiffusionDB and 70 for Civiverse, highlighting potential variation in user interests and content generation across the two platforms. The designated topics were manually defined after reviewing the results of the topic modeling process, ensuring a representative and focused interpretation of the thematic outputs.

DiffusionDB places a stronger emphasis on artistic themes, with a focus on traditional and contemporary art forms, as well as detailed explorations of character design and artistic techniques. Topics such as \textit{Artists and Famous Art Figures}, \textit{Painting Techniques and Styles}, as well as \textit{Portraits} are particularly prevalent. This thematic focus, in contrast to Civiverse, suggests a more specialized environment where users delve deeply into particular artistic practices and engage in creative experimentation within those fields. In contrast, Civiverse, though more thematically diverse, is characterized by a stronger emphasis on explicit and sensual content, with topics such as \textit{Female Genitalia}, \textit{Sensual Clothing}, and \textit{Nudity} being disproportionately represented. This suggests that Civiverse's users are more inclined to produce and share adult-oriented material, with the broader range of explicit content potentially stemming from the platform's more permissive nature.

Despite their thematic differences, both platforms demonstrate significant overlap in topics related to visual and aesthetic quality. Specifically, topics such as \textit{Cinematic Lighting and Effects}, \textit{Aesthetics and Beauty}, \textit{Photography and Realism}, and \textit{Intricate and Complex Details} are prominent across both datasets. This shared emphasis on creating visually striking and high-quality content reflects a common aim of artistic refinement and appeal, suggesting that these elements are consistently important to users of TTI models, regardless of the primary thematic focus.

\begin{figure*}[!htbp]
  \centering
  \begin{minipage}[b]{0.49\textwidth}
    \centering
    \includegraphics[width=\linewidth]{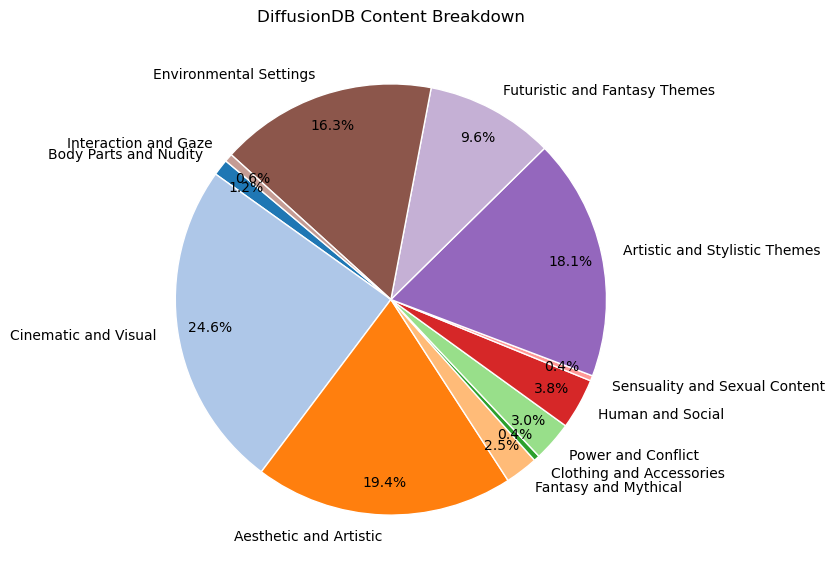}
    \caption{Categorization of the primary topics and their relative proportions within the DiffusionDB dataset.}
    \label{fig:dfdb}
  \end{minipage}%
  \hfill
  \begin{minipage}[b]{0.49\textwidth}
    \centering
    \includegraphics[width=\linewidth]{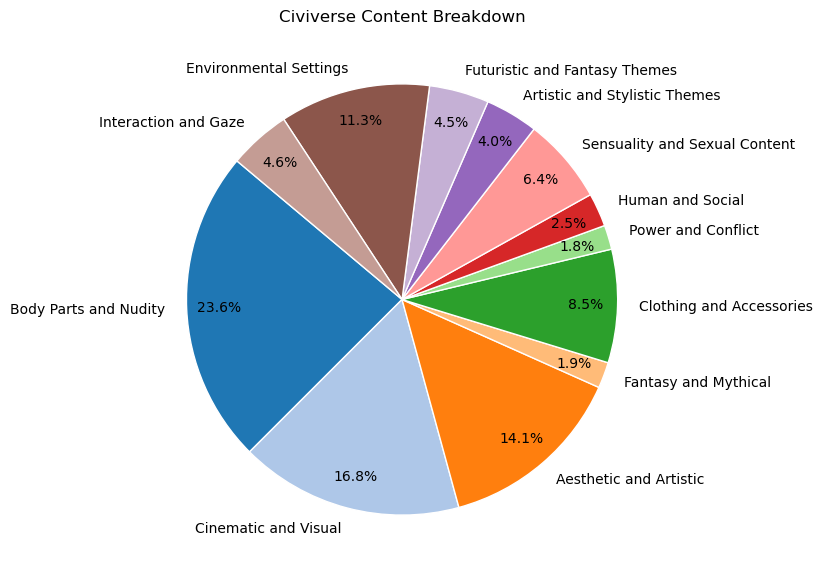}
    \caption{Categorization of the primary topics and their relative proportions within the Civiverse dataset.}
    \label{fig:civiverse}
  \end{minipage}
\end{figure*}

\subsubsection{Originality Scores} 

The results of the analysis reveal distinct differences in originality between the Civiverse and DiffusionDB datasets, with notable patterns emerging across all three originality metrics: lexical, thematic, and word-sequence originality. As demonstrated in Figure \ref{fig:three_images}, prompts from DiffusionDB exhibit a higher proportion of high lexical originality, indicating that users of this platform tend to employ a more diverse and less predictable vocabulary. In contrast, Civiverse displays a greater prevalence of low \textit{lexical originality}, suggesting that prompts in this dataset frequently rely on common and familiar words. Shifting focus to thematic originality, shown in Figure \ref{fig:three_images}, both datasets display a significant focus on low-originality themes; however, this trend is more pronounced in Civiverse, where nearly 70\% of prompts fall into the low thematic originality category, underscoring the platform's reliance on recurring themes. It is important to mention that the very low proportion of high thematic originality observed in both datasets might be influenced by the fact that the topics measured are based solely on those captured through topic modeling. If a prompt addresses uncommon or niche topics that were not identified by the topic modeling process, it remains unnoticed in the categorization, potentially underestimating the presence of its high thematic originality. Finally, in terms of word-sequence originality (Figure \ref{fig:three_images}), both datasets are largely dominated by moderately original word combinations, although Civiverse exhibits a slight advantage in the proportion of high word-sequence originality prompts. 
Overall, the results indicate that while there are instances of originality in user prompts, both datasets exhibit a strong tendency toward the repetition of common topics and familiar word combinations. This trend is particularly evident in Civiverse, where low levels of thematic and lexical originality are more pronounced, suggesting that users on this platform are more inclined to rely on conventional language patterns and familiar topics. Although there are examples of users experimenting with more creative language and concepts, especially in DiffusionDB, where high lexical and word-sequence originality are more prominent, the majority of prompts across both platforms tend to follow established linguistic and thematic structures. To complement this analysis, Table \ref{tab:example_prompts} provides examples of prompts from the DiffusionDB and Civiverse datasets, labeled with their corresponding originality scores, offering a practical illustration of the differences in prompt structures across the two platforms.

\begin{figure}[h!]
    \centering
    \begin{minipage}{0.33\textwidth}
        \centering
        \includegraphics[width=\textwidth]{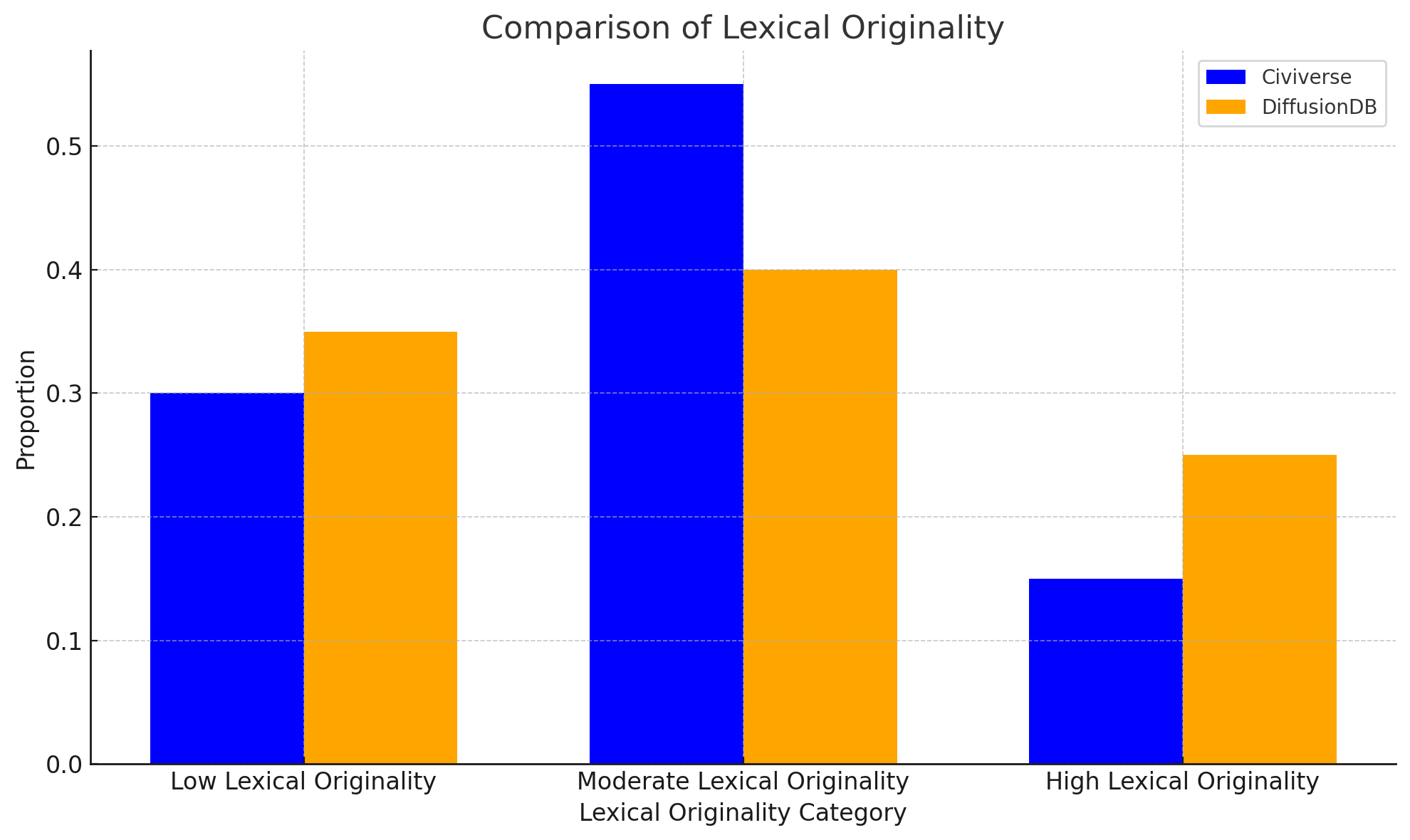}
        \caption*{Lexical Originality}
        \label{fig:lex_orig}
    \end{minipage}\hfill
    \begin{minipage}{0.33\textwidth}
        \centering
        \includegraphics[width=\textwidth]{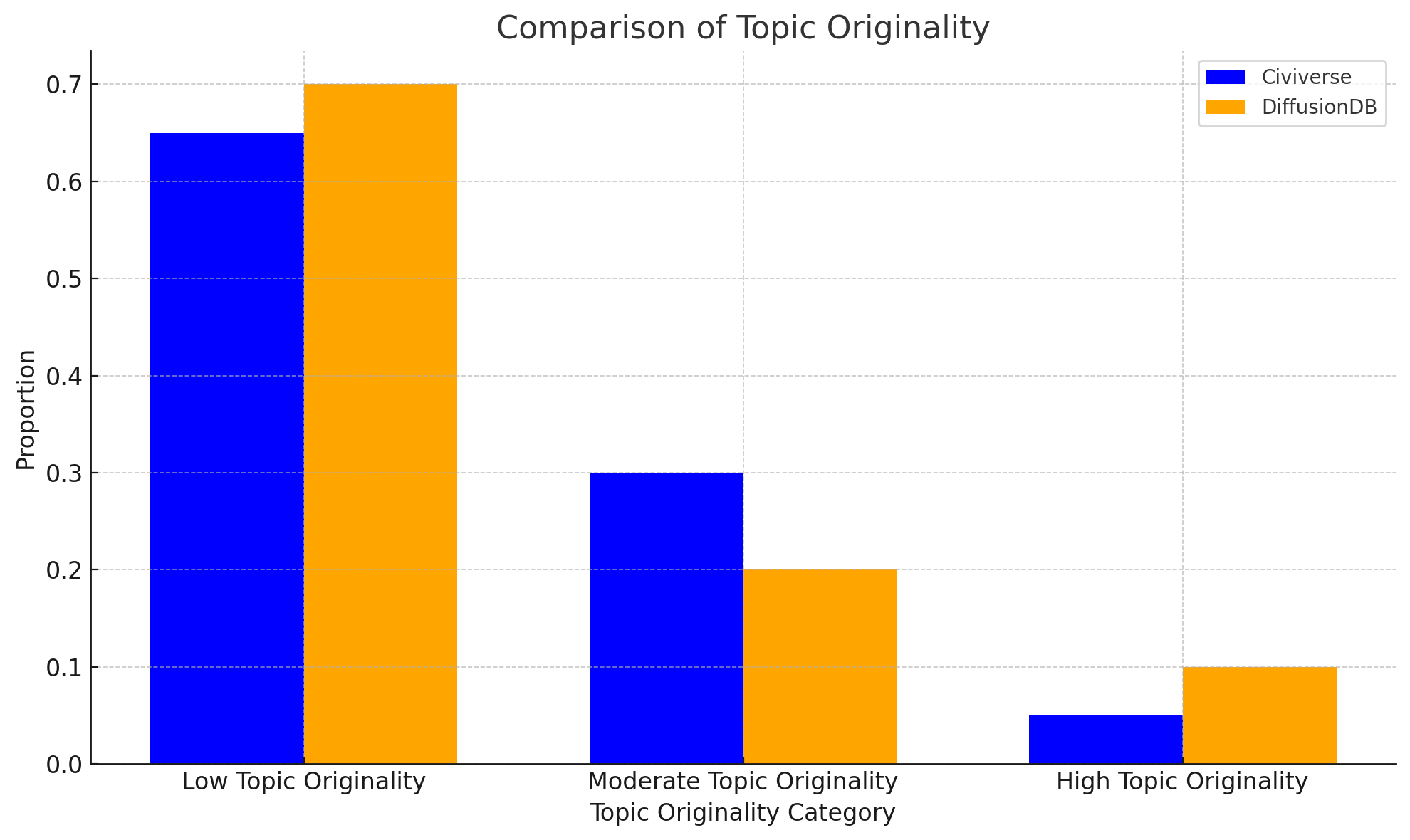}
        \caption*{Thematic Originality}
        \label{fig:topic_orig}
    \end{minipage}\hfill
    \begin{minipage}{0.33\textwidth}
        \centering
        \includegraphics[width=\textwidth]{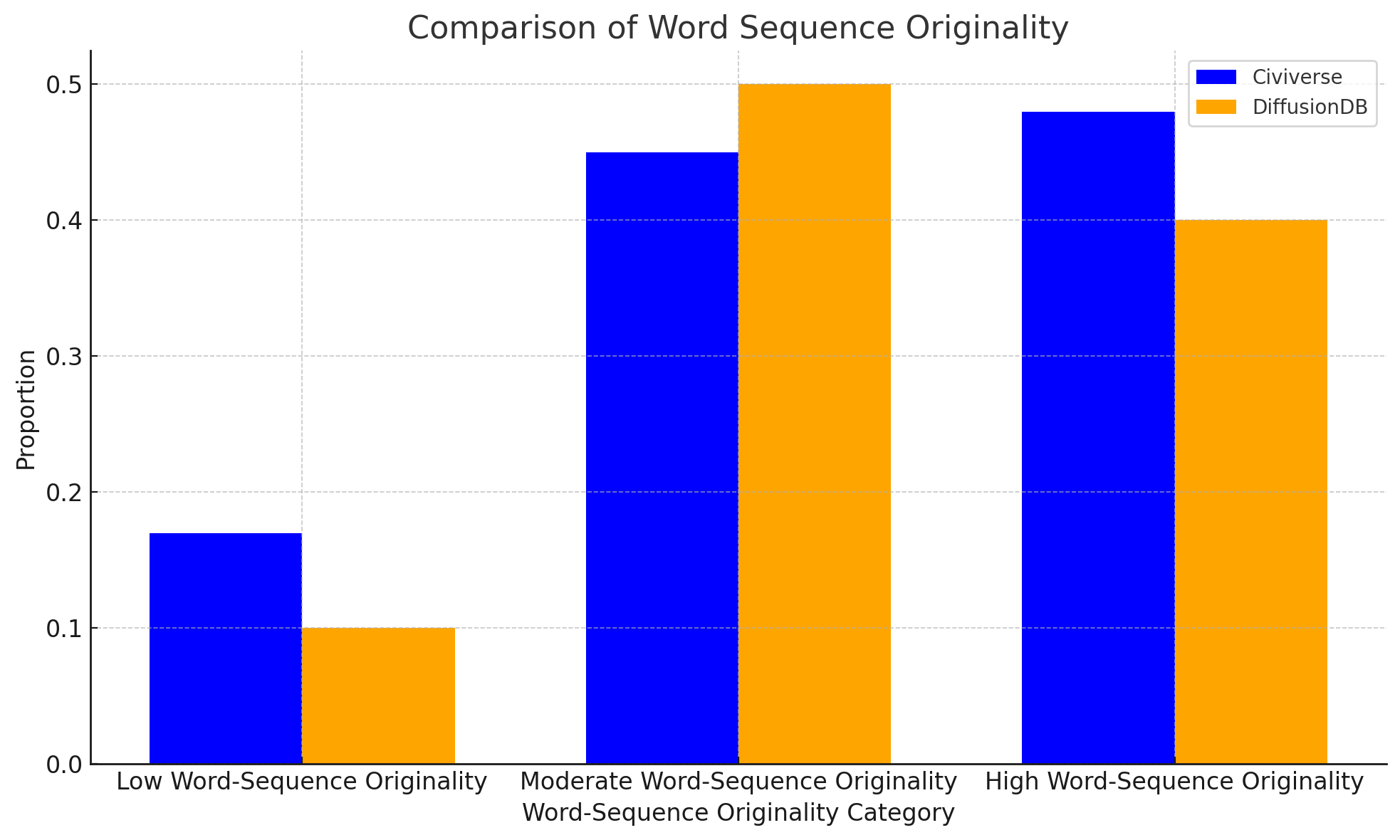}
        \caption*{Word-Sequence Originality}
        \label{fig:word_seq_or}
    \end{minipage}
    
    \caption{Comparison of Lexical, Word-Sequence, and Thematic Originality between DiffusionDB and Civiverse Datasets.}
    \label{fig:three_images}
\end{figure}

\label{fig:three_images}

\clearpage
\begin{table*}[htbp]
\centering
\caption{Example Prompts from DiffusionDB and Civiverse with Assigned Originality Labels.}
\label{tab:example_prompts} 
\renewcommand{\arraystretch}{1.2} 
\setlength{\tabcolsep}{4pt} % Reduced column padding to fit more text
\begin{tabular}{|p{3.2cm}|p{5.5cm}|p{5.5cm}|}
\hline
\textbf{Originality Metrics} & \multicolumn{2}{c|}{\textbf{Prompt Examples}} \\ \hline
                            & \textbf{DiffusionDB Prompt} & \textbf{Civiverse Prompt} \\ \hline
\textbf{Lexical Originality} &                             &                          \\ \hline
Low Lexical Originality      & \textit{a space girl with big and cute eyes, holding a cat, very anime, fine face, realistic shaded perfect face, fine details.}  & \textit{a girl confessing her love, smiling at viewer, blushing, masterpiece, best quality, highres, best illumination, depth of field, detailed background, dynamic angle perspective.}                        \\ \hline
Moderate Lexical Originality & \textit{one cute komodo dragon with very big eyes, wearing a bandana and chain, holding a laser gun, standing on a desk, digital art, award winning, in the style of the movie zootopia.}  & \textit{dark theme low-key mysticism, fantasy, gothic, horror glowing flora nocturnal creatures magical phenomena or eerie dark settings typical of gothic tales, lush roots with Forest Bathed in shadows, macro lens Sepia filter psychedelic. }                       \\ \hline
High Lexical Originality     & \textit{large colorful balloons with people on rope swings underneath, flying high over the beautiful countryside landscape, professional photography, 80 mm, telephoto lens, realistic, detailed, digital art.} & \textit{breathtakingly beautiful woman standing amidst a field of vibrant colorful flowers, the fragrance of the flowers is heavy in the air mixing with the gentle breeze that whispers through the field creating an idyllic and enchanting atmosphere.}                     \\ \hline
\textbf{Thematic Originality} &                             &                          \\ \hline
Low Thematic Originality     & \textit{a painting of a fully dressed girl wearing a jacket upper body with beautiful purple galaxy eyes, highly detailed, digital painting, artstation, sharp focus, dreamy illustration.}   & \textit{gothic style art of two stunningly beautiful naked 20-year-old girls, turquoise eyes, light auburn crown, braid hair, playing with jewelry with a flirtatious smile.}                        \\ \hline
Moderate Thematic Originality & \textit{a painting of an orange cat staring profoundly into the window, american scene painting, dutch golden age.}  & \textit{36-year-old cyberpunk girl, dark fairy tale gown with hand-embroidered details, full skirt and matching accessories, neon-lit rain-drenched alleyway with reflections creating a disorienting effect.   }                    \\ \hline
High Thematic Originality    & \textit{two spherical glass plasma lamp heads with normal human bodies and clothes having an awkward dinner date in a dimly lit cafe.} & \textit{ultra realistic, full body color pencil drawn portrait of Eleanor Rigby, picks up the rice in the church where a wedding has been, lives in a dream, waits at the window, ethereal figures from forgotten memories hover behind her.}                         \\ \hline
\textbf{Word-Sequence Originality} &                        &                          \\ \hline
Low Word-Sequence Originality & \textit{epic gorilla battle, wlop, concept art, digital painting, trending on artstation, highly detailed, epic composition, 8k uhd.}  & \textit{high quality, detailed background, 1girl solo, pink hair, short hair short, losed eyes, looking at viewer, mature female, small breasts, bikini under clothes, wet shirt.}                       \\ \hline
Moderate Word-Sequence Originality & \textit{the subtle shades of consciousness as an abstract painting.}  & \textit{an eldritch magical tome bound in demon skin and an elder sign on the cover.}                       \\ \hline
High Word-Sequence Originality & \textit{a portal to the void deep below the mariana trench, fishes swimming towards the portal, eerie, mixed media. } & \textit{lisbon street, a model in a beaded flowers intertwining, trimmed by lime pink shoelaces through the mesh romper, the A-line romper is made of  white mesh.}                        \\ \hline
\end{tabular}
\end{table*}

\clearpage
\subsection{Case Study: The Impact of Lexical Originality on Visual Homogenization}

In this case study, the impact of \textit{lexical originality} on visual homogenization is explored by analyzing a random sample of 15,000 images and their corresponding prompts, drawn from the \textbf {DiffusionDB} dataset. After the thematic originality metrics are calculated, the prompts are categorized based on their respective topic labels, providing a framework for deeper analysis. To maintain consistency and reduce variability introduced by dissimilar subjects, the focus is then narrowed to prompts specifically related to the topic of \textit{Portraits}. This approach ensures that visual features are compared within a more homogenous subject matter, allowing for more accurate insights into how originality influences visual diversity. As a result, approximately 3,500 prompts and their corresponding images, all labeled under the topic of \textit{Portraits}, are isolated for further investigation and analysis.

\subsubsection{Calculation of CLIP embeddings}
The lexical originality scores for the aforementioned prompts are calculated and categorized into three groups: high originality (975), low originality (1,700), and moderate originality for the remainder ones. To ensure a balanced comparison between the extremes of originality, an equal number of high and low originality cases (975 each) is selected, thereby avoiding bias from unequal representation across originality levels. Using this balanced sample, CLIP embeddings, a widely-used method for mapping both images and text into a shared latent space, are computed for both the corresponding images and text prompts. This enables a quantitative analysis of not only visual, but also textual similarity. By comparing the embeddings of both images and prompts generated from different descriptions, this approach allows for a direct examination of whether lexically similar prompts result in more homogeneous visual and textual outputs.

\subsubsection{Dimensionality Reduction and Clustering}

\begin{figure}[h!]
    \centering
    \begin{minipage}[t]{0.48\textwidth}  % Slightly increase the width to 48% of the text width
        \centering
        \includegraphics[width=\textwidth]{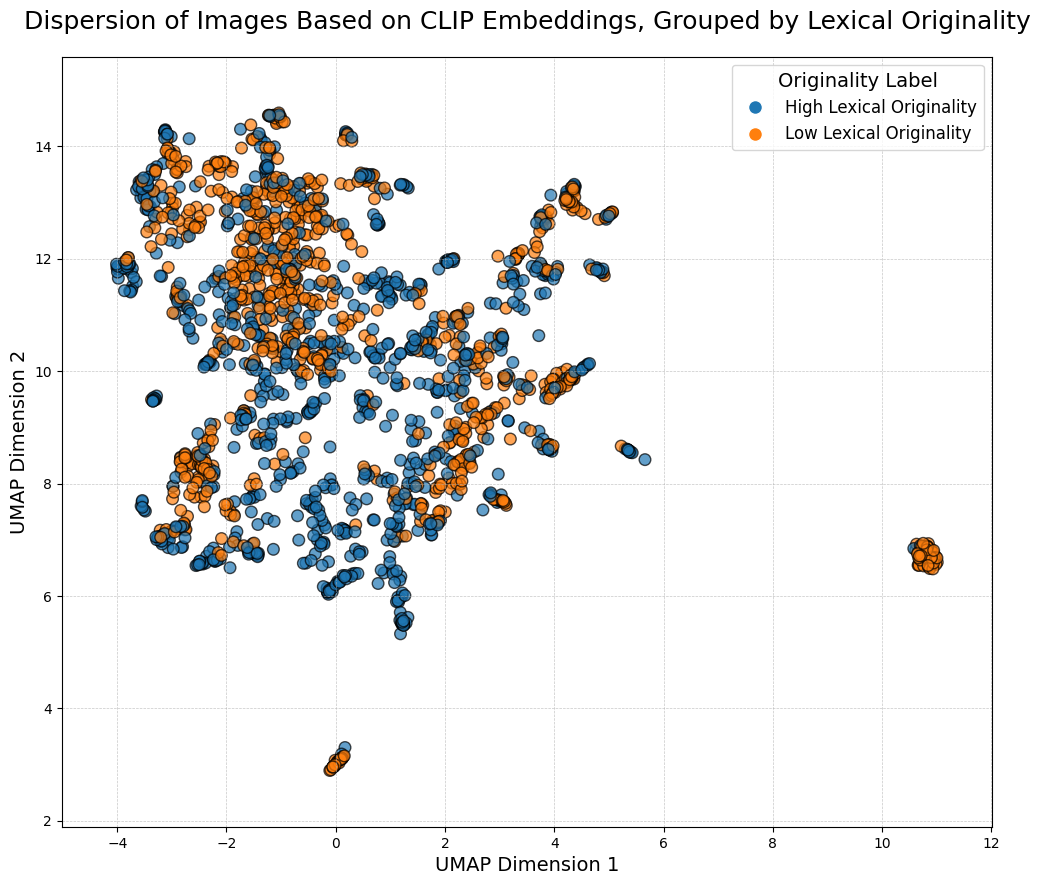}
        \caption{UMAP-based visualization of image clusters from CLIP embeddings.}
        \label{fig:clip_embeddings_images}
    \end{minipage}\hfill
    \begin{minipage}[t]{0.48\textwidth}  % Increase this width as well for symmetry
        \centering
        \includegraphics[width=\textwidth]{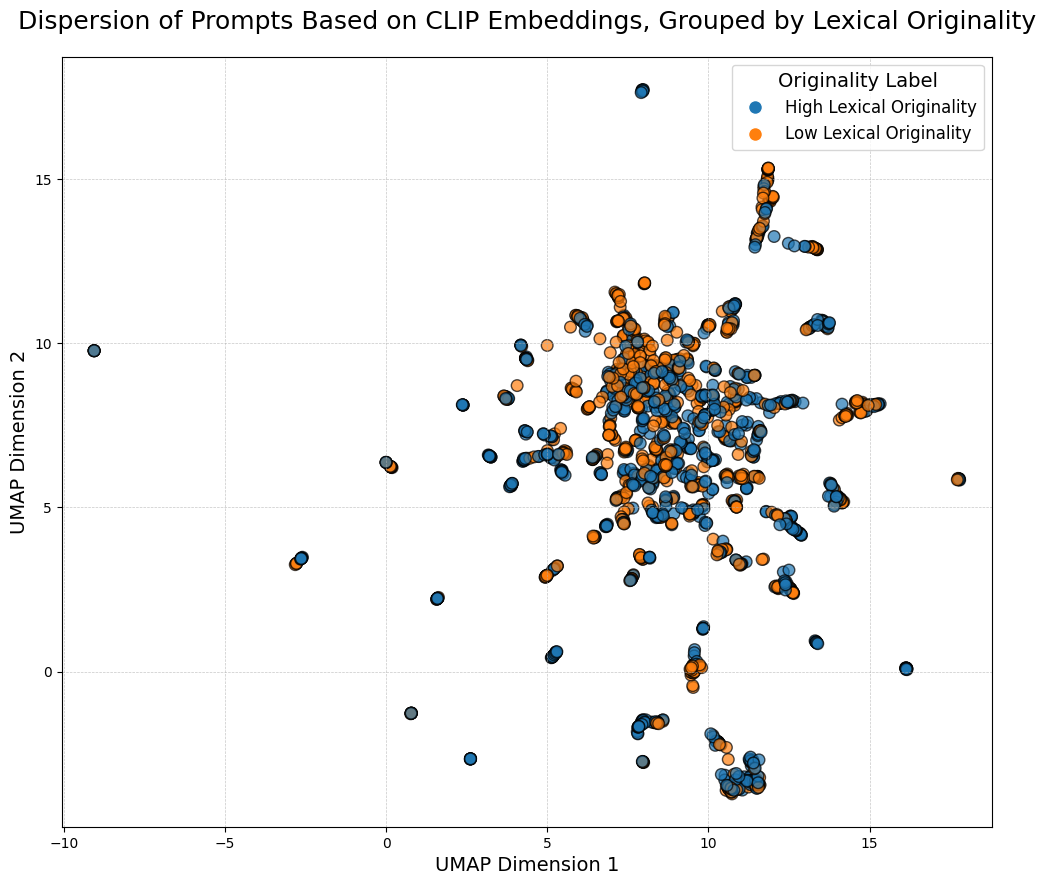}
        \caption{UMAP-based visualization of text prompt clusters from CLIP embeddings.}
        \label{fig:clip_embeddings_text}
    \end{minipage}
    
    \label{fig:embedding_comparison}
\end{figure}

To visualize the clustering of images based on CLIP embeddings, the dimensionality of the high-dimensional embeddings is first reduced using UMAP (Uniform Manifold Approximation and Projection) \cite{mcinnes2018umap}, allowing for a more interpretable two-dimensional representation. The resulting clusters are plotted in scatter plots (see Figures \ref{fig:clip_embeddings_images} and \ref{fig:clip_embeddings_text}), where each point represents an image or a prompt respectively, with colors assigned based on their \textit{lexical originality} labels—either high or low. By overlaying the cluster labels onto these visual representations, distinct patterns of grouping emerge, indicating that both images and prompts with higher lexical originality tend to produce more diverse and dispersed clusters, while lower originality prompts and their corresponding images generate more homogeneous clusters, providing evidence of the relationship between lexical originality and the diversity of both the visual and
textual outputs.

\subsubsection{Quantitative Analysis of Visual Diversity}
To strengthen the analysis, a set of metrics is calculated to assess the dispersion and clustering behavior of of both the generated images and the textual prompts:

\begin{enumerate} \item \textbf{Cluster Dispersion}: The variance of distances within each cluster is used as an indicator of visual diversity, with higher variance reflecting greater variation among the corresponding data points (images or textual prompts) in the cluster.

\item \textbf{Distance from Centroid}: The average distance of each image from the centroid of its cluster is measured to quantify how much individual data points (images or textual prompts) deviate from the cluster's center, with greater distances implying higher diversity. \end{enumerate}

\paragraph{Cluster Dispersion} The variance analysis, conducted for both the visual outputs and the textual CLIP embeddings, reveals that clusters associated with high lexical originality prompts tend to be more dispersed across both dimensions compared to those with low originality. For the images, variances range from 0.313 to 0.852, while the textual embeddings exhibit variances between 0.140 and 0.651.

\paragraph{Average Distance from Centroid} 
Similarly, the average distance from the centroid is found to be larger for both the images and textual prompts generated from high lexical originality prompts, further reinforcing the notion of greater diversity in these outputs. Specifically, for the images, the average distance from the centroid is 6.75, in contrast to 5.42 for those generated from low lexical originality prompts. Likewise, in the textual embedding space, high-originality prompts show an average distance of 5.20, compared to 4.74 for low-originality prompts. 

Overall, the results indicate that lexical originality can be a contributing factor to both visual and textual homogenization, as prompts with lower lexical originality tend to produce more uniform outputs across both domains. This is demonstrated by the tighter clustering, lower variance, and shorter distances from centroids in both the images and textual embeddings generated from low-originality prompts. In contrast, high-originality prompts result in more dispersed clusters and greater diversity, both visually and textually, suggesting that maintaining lexical variation in prompts can be for fostering a broader range of distinct outputs in AI-generated content. 

\begin{figure*}[ht]
    \centering
    % First row of 5 images
    \includegraphics[width=0.19\linewidth]{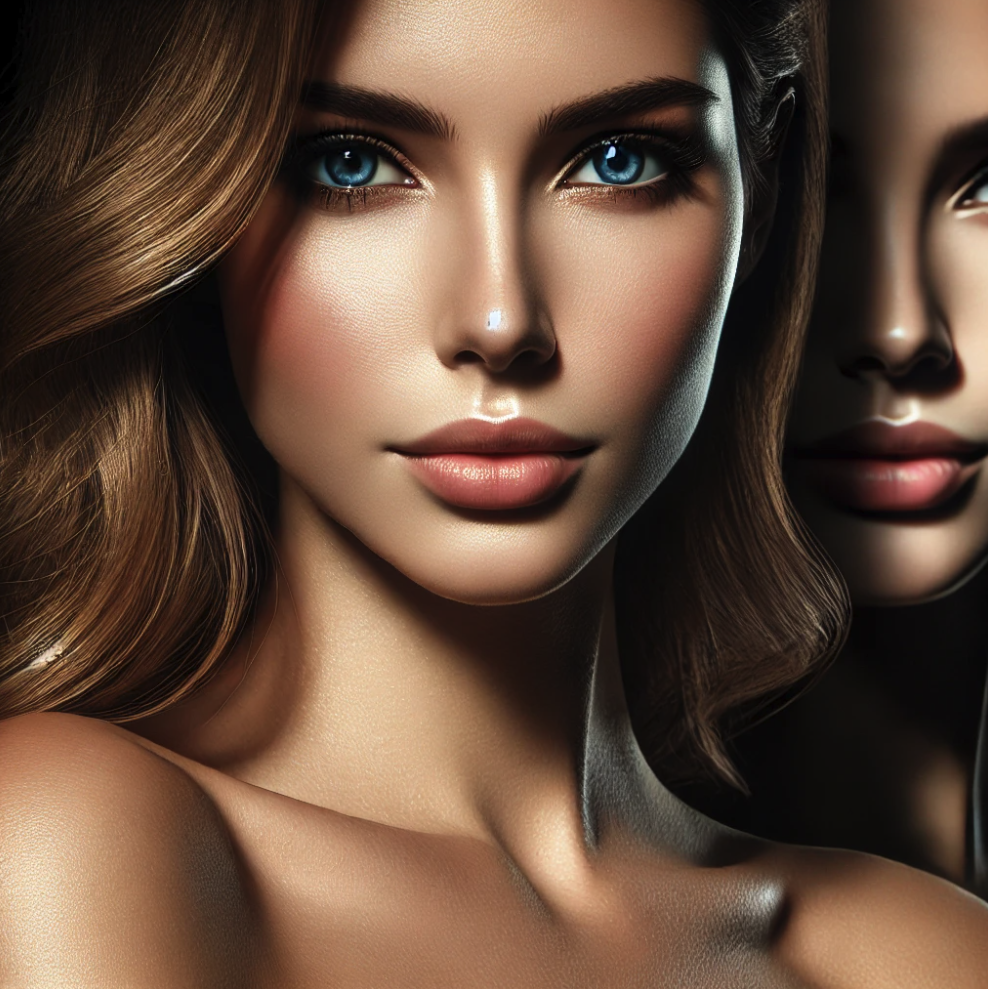}
    \includegraphics[width=0.19\linewidth]{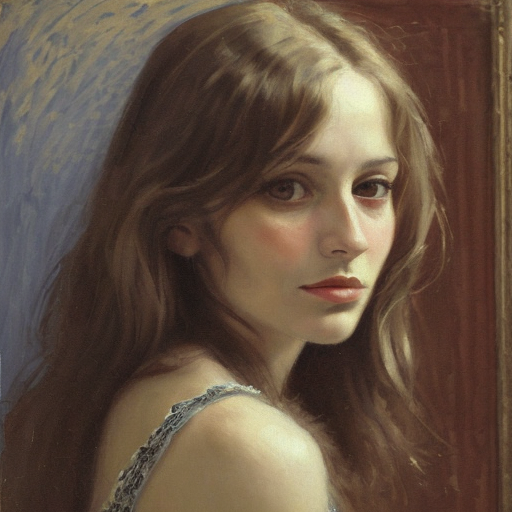}
    \includegraphics[width=0.19\linewidth]{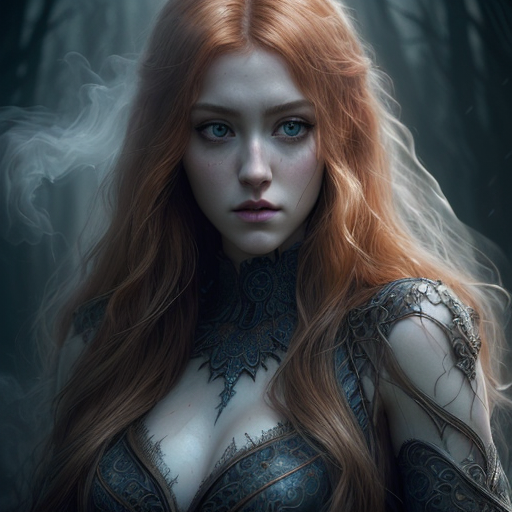}
    \includegraphics[width=0.19\linewidth]{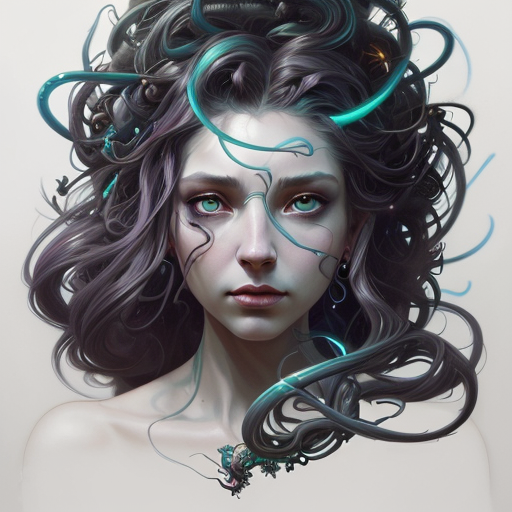}
    \includegraphics[width=0.19\linewidth]{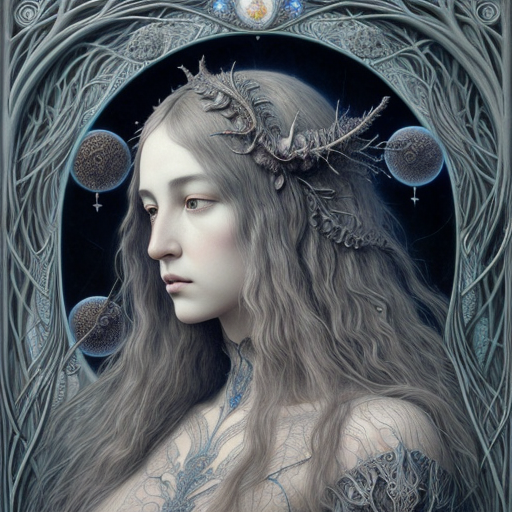}
    
    % Add the prompts as captions beneath each image
    \begin{minipage}{0.19\linewidth}\centering
        \small \textit{"photo, half body portrait of very beautiful woman, realism, extreme detail, real life skin, key art, soft light, volumetric light, 3d shadows photo."} 
    \end{minipage}
    \begin{minipage}{0.19\linewidth}\centering
        \small \textit{"beautiful woman portrait by Jean Leon Gerome"} 
    \end{minipage}
    \begin{minipage}{0.19\linewidth}\centering
        \small \textit{"portrait, highly detailed, red hair woman, pale skin, fantasy setting, forest with mist, soft light, dark mysterious."} 
    \end{minipage}
    \begin{minipage}{0.19\linewidth}\centering
        \small \textit{"a futuristic portrait painting, short dark brown messy haircut, large green eyes, antichrist eyes, slightly rounded face, pointed chin, thin lips, small nose, hd painting, cyberpunk inspiration."} 
    \end{minipage}
    \begin{minipage}{0.19\linewidth}\centering
        \small \textit{"detailed realistic beautiful gothic moon goddess face portrait by jean delville, gustave dore, surreality, hyperdetailed ultrasharp octane render."} 
    \end{minipage}
    \caption{Portrait images generated from prompts low in lexical originality.}
    \label{fig:first_set}
\end{figure*}

\vspace{0.5cm} % Reduced space between two sets of images

\begin{figure*}[ht]
    \centering
    % Second row of 5 images
    \includegraphics[width=0.19\linewidth]{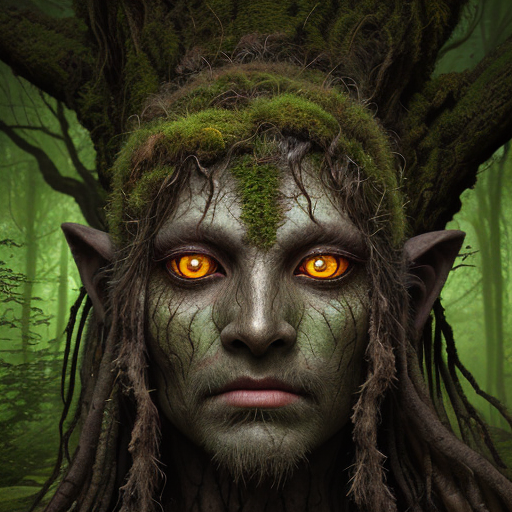}
    \includegraphics[width=0.19\linewidth]{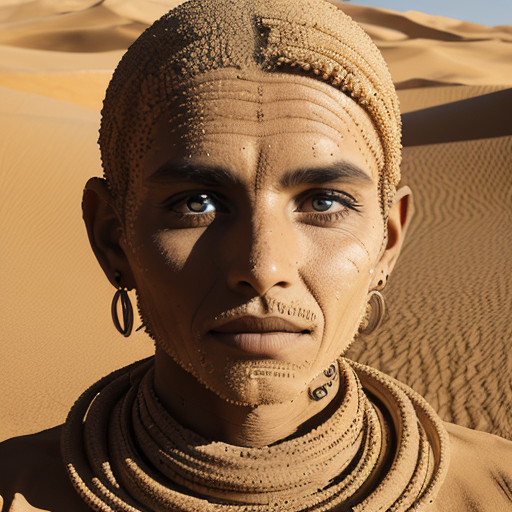}
    \includegraphics[width=0.19\linewidth]{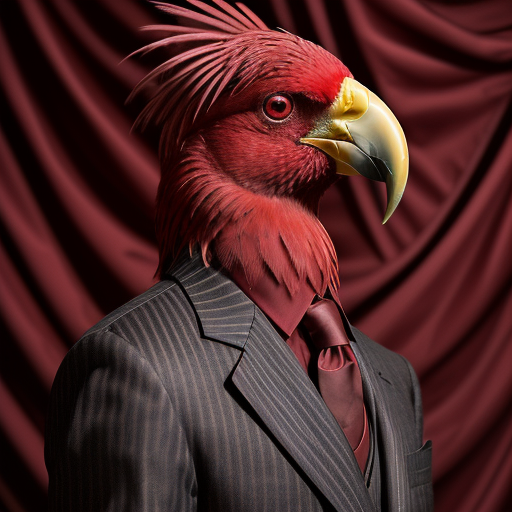}
    \includegraphics[width=0.19\linewidth]{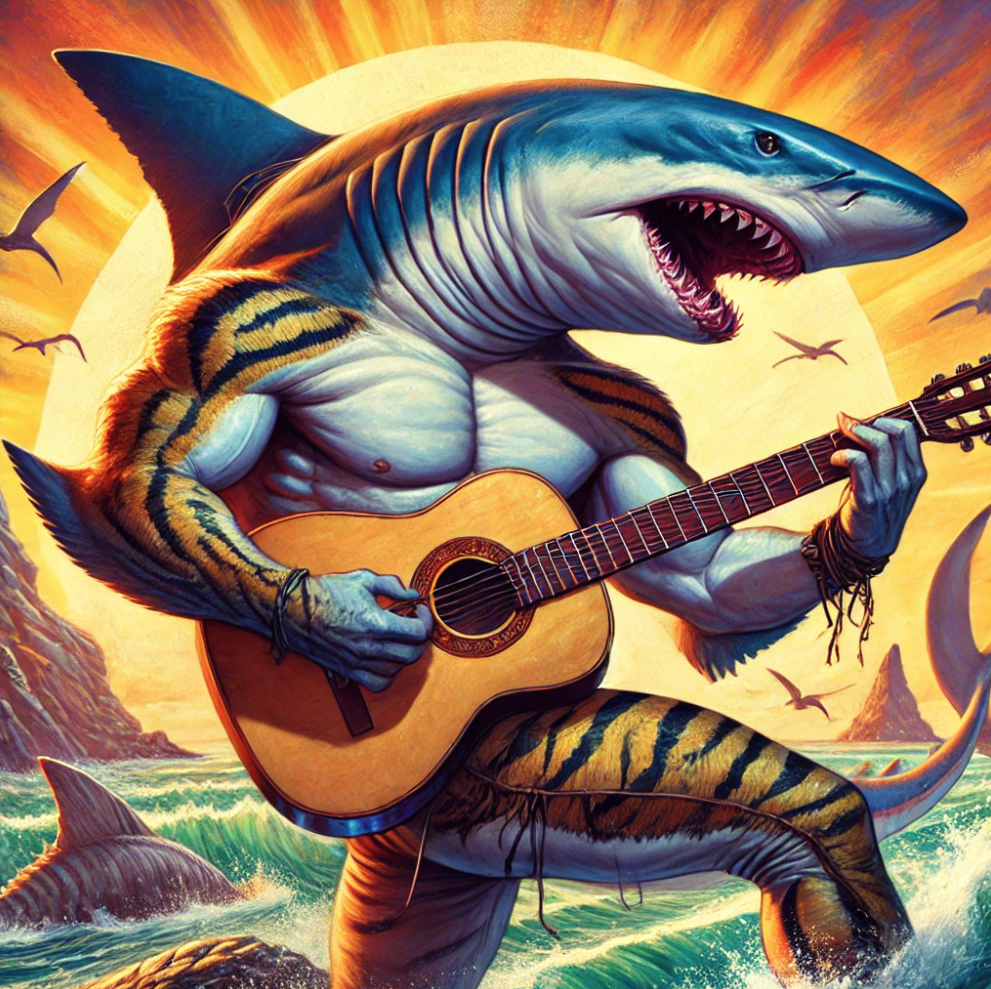}
    \includegraphics[width=0.19\linewidth]{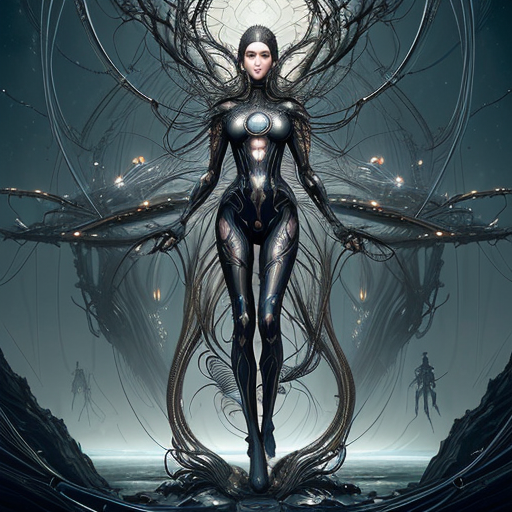}
    
    % Add the prompts as captions beneath each image
    \begin{minipage}{0.19\linewidth}\centering
        \small \textit{"wise forest spirit detailed portrait, bark-textured skin, glowing amber eyes, vines growing from hair, connection with nature."} 
    \end{minipage}
    \begin{minipage}{0.19\linewidth}\centering
        \small \textit{"a desert nomad portait, made entirely of shifting sand."} 
    \end{minipage}
    \begin{minipage}{0.19\linewidth}\centering
        \small \textit{"realistic portrait of a red bird in a business suit, wearing a tie, dramatic red draped fabric backdrop, intelligent, sharp feathers, confident in a formal pose."} 
    \end{minipage}
    \begin{minipage}{0.19\linewidth}\centering
        \small \textit{"portrait of majestic anthropomorphic shark, holding a guitar, late jurassic, alain beneteau, standing on a cliffside, blending warm tones."} 
    \end{minipage}
    \begin{minipage}{0.19\linewidth}\centering
        \small \textit{"full body cyberpunk-themed portrait, futuristic woman, high-tech bodysuit, biomechanical elements, dystopian environment with glowing lights and shadowy figures, metallic textures, emerging from the water."} 
    \end{minipage}
    
    \caption{Portrait images generated from prompts high in lexical originality.}
    \label{fig:second_set}
\end{figure*}

\subsection{{The Impact of Prompt Characteristics on User Engagement}}

\subsubsection{Model Performance Metrics}

The performance metrics of the predictive model, which estimates user engagement in the form of \textit{LikeCount}, indicate both a strong fit and reasonable accuracy. The model, specifically designed to predict the popularity of an image based on various prompt characteristics, is tailored for the Civiverse platform, where \textit{LikeCount} serves as a key indicator of user interaction and engagement. With an R-squared (R²) value of 0.8266, the model successfully explains approximately 82.66\% of the variance in user engagement, capturing potential underlying patterns that can drive an image's visual popularity.

Additionally, the Mean Absolute Error (MAE) of 5.6015 indicates that, on average, the model's predictions deviate from the actual number of likes by about 6 units. This demonstrates that the model is relatively accurate in estimating user engagement, providing valuable insights into how different prompt characteristics influence the popularity of AI-generated images. 

\begin{table}[h!]
\centering
\caption{Model Performance Metrics}
\begin{tabular}{|l|r|}
\hline
\textbf{Metric}               & \textbf{Value} \\ \hline
R-squared (R²)                & 0.8266        \\ \hline
Mean Absolute Error (MAE)      & 5.6015        \\ \hline
Root Mean Squared Error (RMSE) & 6.1683        \\ \hline
\end{tabular}
\end{table}

\subsubsection{Feature Importance}

A closer examination of the feature coefficients, presented in Table \ref{tab:feature_importance}, reveals several key factors that strongly influence user engagement on the CivitAI platform, with certain thematic and visual elements, rather than originality metrics, being the most influential drivers of engagement.

Among the factors driving user engagement, content featuring \textit{feminine subjects} emerges as the most significant predictor, with a coefficient of 4.4282. This indicates a pronounced user preference for feminine aesthetics, a trend that is further reinforced by the strategic use of \textit{charismatic adjectives}. With a coefficient of 3.2054, these adjectives underscore the importance of enhancing beauty and appeal, suggesting that users are drawn to content that emphasizes visual attractiveness. Additionally, the inclusion of dynamic visual elements, such as varied \textit{angles and perspectives}, significantly boosts engagement, as evidenced by a coefficient of 3.7247. This indicates that users are particularly drawn to content that feels more interactive and engaging, as if the images invite the viewer into the scene. Moreover, fantasy elements also contribute notably to user engagement, with a coefficient of 3.3014, highlighting users' interest in visually imaginative themes. Finally, the analysis reveals that explicit anatomical details, such as the representation of breasts, play a role in attracting user attention, as reflected by the coefficient of 2.7839, indicating that users are drawn to content with nudity and intimate visual focus.

\begin{figure*}[ht]
    \centering
    \includegraphics[width=0.9\linewidth]{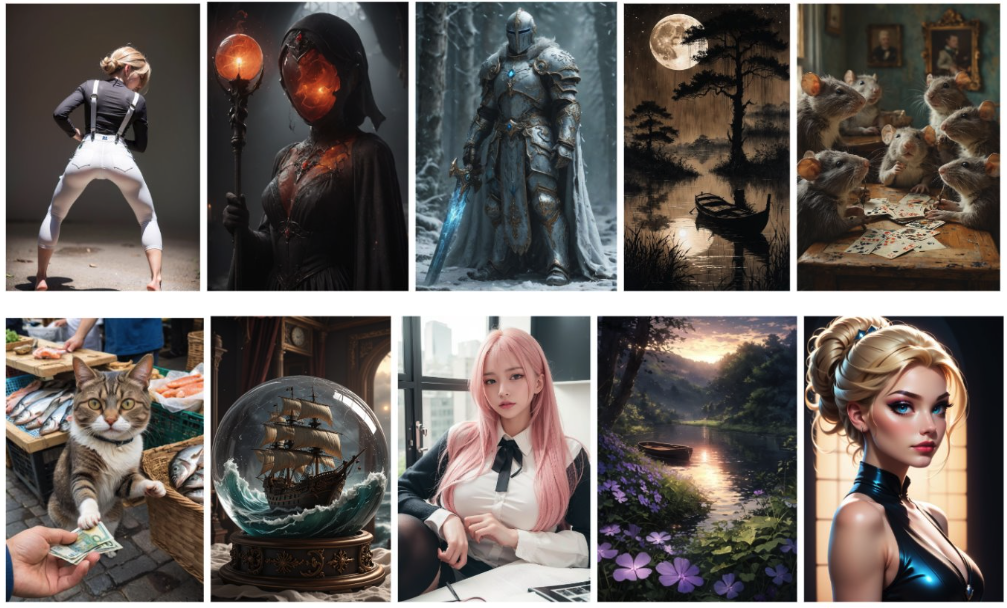} % Adjust the width to make it bigger
    \caption{Most popular images in the Civiverse dataset as of May 6th, 2024.}
    \label{fig:popular_images}
\end{figure*}

Interestingly, these findings align with some of the most popular images in the Civiverse dataset, where user engagement mirrors the trends identified in the feature analysis. As seen in Figure \ref{fig:popular_images}, the ten most popular images on the platform as of May 6th, 2024, prominently feature themes such as fantasy, charismatic feminine subjects, and visually dynamic compositions. 

\begin{table}[h!]
\centering
\caption{Feature Coefficients and Statistical Significance}
\label{tab:feature_importance}
\begin{tabular}{lcccc}
\toprule
\textbf{Feature} & \textbf{Coefficient} & \textbf{Std. Error} & \textbf{p-value} & \textbf{Confidence Interval} \\ 
\midrule
Feminine Subjects & 4.4282 & 0.0293 & 0.0026 & [4.3708, 4.4856] \\
Angles, Perspectives, and Viewer Interaction & 3.7247 & 0.0439 & 0.0082 & [3.6386, 3.8108] \\
Fantasy Elements & 3.3014 & 0.0293 & 0.0255 & [3.2440, 3.3588] \\
Charismatic Adjectives & 3.2054 & 0.0365 & 0.0404 & [3.1338, 3.2770] \\
Breast Details and Chest Anatomy & 2.7839 & 0.0305 & 0.0099 & [2.7241, 2.8438] \\
\bottomrule
\end{tabular}
\end{table}

\section{Discussion}

The findings of this study emphasize that human agency, particularly through user-generated text descriptions or prompts, plays a pivotal role in shaping AI-generated visual culture. This influence extends beyond the dominant patterns learned by algorithms from their training data, as it is also deeply rooted in user behavior, perpetuating biases, particularly in relation to gender representation, and contributing to the homogenization of visual content.

\subsection{Surface Aesthetics and Prompting}

Although both DiffusionDB and Civiverse exhibit a considerable overlap in their focus on visual and aesthetic quality, the strategies employed by users to achieve this are often formulaic and lack creative diversity. Across both platforms, users tend to rely on a predictable set of prompt keywords, such as \textit{cinematic}, \textit{highly detailed}, or \textit{8K} resolution, to produce visually striking outputs. However, this reliance on standardized terms reveals a broader tendency to prioritize surface-level aesthetics over more diverse, experimental, or innovative approaches to image generation.

While the prompts generated in the DiffusionDB dataset emphasize artistic themes, particularly focusing on traditional and contemporary art forms, users still frequently rely on mimicking established styles, often incorporating phrases like \textit{in the style of} or \textit{art by}. This reliance on familiar artistic frameworks and visual aesthetics indicates that even in a context where users are engaged with artistic themes,  there is a tendency to adhere to established styles rather than exploring unconventional or innovative approaches. In contrast, prompts from the Civiverse dataset reveal a pronounced focus on explicit content, a trend that can be explained through the platform’s more permissive content moderation policies. Consequently, the overrepresentation of such content underscores a prioritization of sensationalism over deeper creative exploration, leading to a homogenized visual culture that is narrowly focused on themes related to gender and the human body.

\subsection{The Impact of User Interaction}

\paragraph{Visual Homogenization} 
The analysis conducted on both DiffusionDB and Civiverse datasets reveals a consistent user tendency to rely on less original prompts, largely influenced by platform trends and guidelines. The reliance on familiar keywords and stylistic frameworks creates a feedback loop that shapes the generated visuals, resulting in a homogeneous and repetitive visual landscape. This overuse of similar subjects, artistic techniques, and color palettes not only contributes to visual monotony but also stifles creative diversity, as prompts with limited thematic exploration and unconventional word combinations shift the focus from innovation to the production of formulaic, visually appealing content. This uniformity not only restricts opportunities for experimentation and the blending of new ideas, but also raises broader concerns about authorship in AI-generated content. Specifically, as familiar visual elements are continually reproduced, the machine begins to appear as the primary creative agent, further complicating the question of authorship in the creative process.

\paragraph{Perpetuation of Gender Bias}
The analysis of user engagement on the Civiverse platform reveals that aesthetic appeal factors, such as the portrayal of feminine subjects, the use of charismatic adjectives, and dynamic visual elements, significantly impact image popularity more than prompt originality, as assessed by the proposed originality metrics. This observation is reinforced by the results from the user engagement prediction model, which reveals that the most popular AI-generated images on the platform often feature female subjects described with adjectives such as \textit{beautiful}, \textit{attractive}, or \textit{seductive}. These images frequently depict women in sexualized poses or interactions designed to appeal directly to viewers, such as making eye contact. This pattern underscores a user-driven preference for content that prioritizes surface-level aesthetics, often narrowing female representation to a limited and objectified form of beauty. Such trends not only reflect broader cultural norms regarding aesthetic ideals but also demonstrate how user input and content popularity perpetuate specific biases, particularly concerning gender, in AI-generated visual content.

\section{Limitations and Future Work} 

While this study provides valuable insights into the role of user behavior in shaping AI-generated content, several limitations must be acknowledged. First, the datasets used, DiffusionDB and Civiverse, are platform-specific, reflecting the characteristics of their respective user communities and content moderation policies. As a result, the findings may not be generalizable to other platforms with different user bases or guidelines.

Second, while the originality metrics developed in this study do offer a novel way to assess prompt originality by capturing various aspects of word usage and thematic combinations, they do not fully encompass the broader, more nuanced dimensions of human expression. Specifically, lexical originality measures word uniqueness but overlooks the semantic richness or conceptual depth a prompt might convey, meaning prompts with common words could still reflect innovative ideas yet score low in originality. Similarly, while thematic originality measures the rarity of topics, it doesn’t account for how innovative or coherent the combinations are, allowing prompts with rare but incoherent themes to score highly. Additionally, word-sequence originality, though useful for identifying uncommon word pairings, does not evaluate whether these combinations contribute to a coherent narrative or visual storytelling. Thus, while these metrics offer valuable insights into linguistic originality by examining prompts from a lexical and thematic pattern perspective, assessing how similar or distinct the linguistic choices are among users when interacting with TTI models, they may overlook more complex aspects of human expression, such as conceptual innovation or narrative coherence, which go beyond simple word usage and thematic combinations.

Future research could integrate a "human-in-the-loop" approach, where human evaluators assess the originality not only of the prompts but also of the images generated from them. This would involve having human assessors rate the creativity, novelty, and artistic value of the outputs, offering subjective insights that go beyond what can be measured by originality metrics alone. By comparing human evaluations with the numerical originality scores, one could better understand whether prompts with higher originality metrics actually result in more diverse and visually compelling images, or if the metrics overlook important elements of human-perceived innovation and conceptual richness.

\section{Conclusion}

In this study, we examined the critical influence of user behavior on the diversity of AI-generated content, particularly focusing on the role of user generated prompts on visual homogenization. Through an in-depth analysis of lexical, thematic, and word-sequence originality in the DiffusionDB and Civiverse datasets, we uncovered that low prompt originality plays a significant role in driving the creation of repetitive and visually uniform content. Our findings highlight that this homogenization is not merely a consequence of the training data alone but is also related to users' reliance on standardized, formulaic prompting practices, which in turn reinforce visual uniformity and limit creative exploration. However, when users move beyond established templates and incorporate greater creative variability into their prompts, we observed that the resulting AI-generated images demonstrated a higher degree of visual diversity, underscoring the pivotal role of user input in shaping the originality and richness of AI-generated outputs.

Beyond visual uniformity, our research also delved into the impact of user input on the reinforcement of cultural biases, particularly those related to gender representation. By examining feature importance in the predictive linear regression model of user engagement, we found that popular images on the Civiverse platform often reflect user preferences for aesthetically appealing content that emphasizes surface-level traits such as beauty, dynamic compositions, and explicit themes. This tendency, especially in the repeated portrayal of feminine subjects in narrow, objectified forms, highlights how user behavior shapes not only the aesthetic qualities of AI-generated images but also perpetuates existing cultural biases within these visual outputs.

Ultimately, this study emphasizes the need for platform developers and online communities to reconsider prompt engineering guidelines and foster environments that encourage user exploration when interacting with AI content generation tools. The aforementioned findings point to the broader implications of human-AI collaboration, advocating for active user involvement to prevent the erosion of creativity and promote a richer, more diverse AI-generated visual culture.

\section{Acknowledgements}
This work was supported by Swiss National Science Foundation (Ambizione
Grant 216104).

\bibliographystyle{splncs04}
\bibliography{main}
\end{document}